\documentclass[review]{elsarticle}
\usepackage{geometry}
\usepackage[usenames,dvipsnames]{color}
\usepackage{lineno}
\modulolinenumbers[5]
\usepackage{graphicx}
\usepackage{booktabs} 
\usepackage{tabularx}
\usepackage{textcomp}
\usepackage{graphicx}
\usepackage{caption}
\usepackage{subcaption}
\usepackage{amssymb}
\usepackage{amsmath,amssymb,amsfonts}
\usepackage[toc,page]{appendix}
\usepackage[autostyle]{csquotes}
\usepackage{array,longtable,ltxtable,filecontents}
\usepackage{array, caption, tabularx,  ragged2e,  booktabs}

\usepackage[ruled,vlined]{algorithm2e}
\usepackage{textcomp}
\usepackage{multirow}
\usepackage{xcolor}
\usepackage{adjustbox}
\usepackage{caption}    
\usepackage{booktabs}
\usepackage[T1]{fontenc}
\usepackage[font=small,labelfont=bf,tableposition=top]{caption}
\usepackage[bookmarks,bookmarksnumbered]{hyperref}
\hypersetup{colorlinks = true,linkcolor = blue,anchorcolor =red,citecolor = blue,filecolor = red,urlcolor = red, pdfauthor=author}

\let\today\relax
\makeatletter
\def\ps@pprintTitle{%
    \let\@oddhead\@empty
    \let\@evenhead\@empty
    \def\@oddfoot{\footnotesize\itshape
         {} \hfill\today}%
    \let\@evenfoot\@oddfoot
    }
\makeatother

\DeclareCaptionLabelFormat{andtable}{#1~#2  \&  \tablename~\thetable}

\usepackage{blindtext}
\usepackage{array,hhline}

\usepackage{array}
\newcolumntype{P}[1]{>{\centering\arraybackslash}p{#1}}
\newcolumntype{M}[1]{>{\centering\arraybackslash}m{#1}}

\journal{Expert Systems with Applications}

\biboptions{authoryear}

\begin{document}

\begin{frontmatter}

\title{Look Before You Leap: Detecting Phishing Web Pages by Exploiting Raw URL And HTML Characteristics}


\author{Chidimma Opara\corref{cor1}\fnref{label1}}
\author{Yingke Chen\fnref{label2}}
\author{Bo Wei\fnref{label3}}
 \fntext[label1]{Teesside University, TS1 3BX, Middlesbrough, UK}
 \fntext[label2]{Northumbria University, NE1 8QH, Newcastle upon Tyne, UK }
 \fntext[label3]{Newcastle University, Newcastle upon Tyne, NE1 7RU, UK }

\begin{abstract} 
Phishing websites distribute unsolicited content and are frequently used to commit email and internet fraud; detecting them before any user information is submitted is critical. Several efforts have been made to detect these phishing websites in recent years. Most existing approaches use hand-crafted lexical and statistical features from a website's textual content to train classification models to detect phishing web pages. However, these phishing detection approaches have a few challenges, including 1) the tediousness of extracting hand-crafted features, which require specialized domain knowledge to determine which features are useful for a particular platform; and 2) the difficulties encountered by models built on hand-crafted features to capture the semantic patterns in words and characters in URL and HTML content. To address these challenges, this paper proposes WebPhish, an end-to-end deep neural network trained using embedded raw URLs and HTML content to detect website phishing attacks. First, the proposed model automatically employs an embedding technique to extract the corresponding characters into homologous dense vectors. Then, the concatenation layer merges the URL and HTML embedding matrices. Following that, Convolutional layers are used to model its semantic dependencies. {Extensive experiments were conducted with real-world phishing data, which yielded an accuracy of 98.1\%}, showing that WebPhish outperforms baseline detection approaches in identifying phishing pages.

\end{abstract}



\end{frontmatter}


\section{Introduction}
Phishing has recently become a preferred method of attack for cybercriminals due to its low cost and limited technical skill requirements. The majority of phishing attacks begin with spam emails. Frequently, these emails contain links to phishing web pages. In April 2020 alone, Gmail intercepted over 100 million spam emails daily, including 18 million COVID-19 pandemic phishing attacks.\footnote{J. Tidy, “Google blocking 18m coronavirus scam emails every day,”
https://www.bbc.co.uk/news/technology-52319093, accessed: 2020-04-20.} Due to the magnitude of this cyber-attack, industrial and academic experts have put much effort into combating phishing and invented various anti-phishing solutions.


Recent research in phishing detection approaches has resulted in the rise of multiple technical methods, such as augmenting password logins \cite{chattaraj2018new} and multi-factor authentication \cite{acar2013single}. However, these techniques are usually server-side systems that require the Internet user to correspond with a remote service, which adds further delay in the communication channel. Another popular phishing detection system that relies on a centralized architecture is the phishing blacklist and whitelist methods \cite{Google}. A URL visited by an internet user will be compared with the URL in these lists in real time. Although the list-based methods tend to keep the false positive rate low, however, a significant shortcoming is that the lists are not exhaustive, and they fail to detect zero-day phishing attacks. To mitigate these limitations, researchers have developed several anti-phishing techniques using machine learning models as they are mostly client-side based and can generalize their predictions on unseen data.

Detecting web page phishing using machine learning typically involves extracting an appropriate feature representation from the web page component and training machine learning-based prediction models on that representation. 
There has been extensive research on lexical features, host-based features, content-based features, and even context and popularity-based features \cite{lecun2015deep, gutierrez2018learning, buber2017detecting}. 
While the strategies outlined above have demonstrated success, they do have some limitations: 1) The inconvenient nature of extracting hand-crafted features, which require specialized domain knowledge to determine which features are useful for a given platform; 2) The need to continuously update the hand-crafted features in order to remain relevant when faced with new phishing techniques. 


Additionally, from our literature review, we found that existing research has commonly studied the performance of models using just raw URLs \cite{le2018urlnet} and \cite{bahnsen2017classifying} or HTML content \cite{9207707} when detecting phishing on web pages using deep learning models. However, based on our research and experiments, we reveal that the information in different parts of a web page can provide different characteristics when detecting phishing. For example, the HTML content of the page gives the semantic and structural characteristics, and the URL can provide insight into leveraging the web page's Internet address. Therefore, leveraging the learned semantic, lexical, and syntactic ambiguities from URL and HTML content to detect phishing is essential.

The objective of this paper is:

\begin{enumerate}
    \item To accurately detect phishing web pages with minimal false positives and false negatives using only the URL and HTML’s raw content on a deep learning model.
    \item To provide a phishing detection model that is not reliant on third-party systems (such as search engines and optical readers) and databases (such as blacklists/ whitelists).
    \item To develop a web page phishing detection model that does not rely on hand-crafted features, which needs extensive domain knowledge.
\end{enumerate}


Consequently, we propose WebPhish, an end-to-end deep neural network model that detects phishing attacks using raw URLs and HTML content. First, we employ an embedding technique to generate homologous dense vectors from the corresponding URL characters and HTML words. The concatenation layer then combines the URL and HTML embedding matrices. Convolutional layers are then used to model the semantic relationships between the words and characters.
Specifically, WebPhish uses an embedding technique to generate homologous dense vectors from the corresponding characters in the URL and words in the HTML. The concatenation layer then combines the URL and HTML embedding matrices. Following that, convolutional layers are used to model the semantic relationships between the words. To enable ease of verification and replicability, we have made the dataset available to other researchers interested in this topic. 

The following are the main contributions of this work:
\begin{itemize}
\item This paper proposes WebPhish, which uses only the raw content of the URL and HTML document of a web page to train a deep neural network model for phishing detection. Manual feature engineering is reduced as WebPhish learns the representation in the features of the HTML document, and we do not depend on any third-party system. Our proposed approach takes advantage of the word and character embedding matrix to present a phishing detection model that automatically accommodates new web content and is, therefore, easily applied to test data.

\item A thorough empirical evaluation based on collected real-world data yields results that show that the proposed model significantly outperforms state-of-the-art methods demonstrating the validity of our approach.


\item Furthermore, we carried out an ablation study on the efficiency of the proposed model on other textual datasets to demonstrate its adaptability and flexibility beyond the proposed domain.

\end{itemize}

The remainder of the paper is organized as follows: the next section provides an overview of related works on proposed techniques for detecting phishing on web pages. Section 3 gives an in-depth description of our proposed model, while Section 4 elaborates on the dataset collection and evaluation metrics used to analyze WebPhish. The detailed results of the evaluations of our proposed model are in Section 5. Section 6 discusses the research conducted and the impact of the DNN in the proposed model. Finally, we conclude our paper in Section 7.

\section{Related Works}
This section surveys the state-of-the-art techniques for detecting phishing web pages using manual feature engineering and automatic feature extraction techniques applied to machine-learning algorithms.

\subsection{Phishing Detection Using Machine Learning Algorithms with Handcrafted Features}
The techniques reviewed in this section are based on the assumption that the infrastructure of phishing pages differs from that of legitimate pages. As a result, extracting specific lexical and statistical features from web pages is expected to differentiate phishing web pages from legitimate ones. The referred features include URLs, HTML, and DNS commonly found on web pages \cite{amrutkar2017detecting}. Recently, NLP-based features such as Bag of Words, ngrams, and TF-IDF have been generated from web page contents to determine their legitimacy \cite{rendall2020towards,lecun2015deep, gutierrez2018learning, buber2017detecting}. In most published state-of-the-art approaches, these extracted features are used by machine learning algorithms to classify a web page as phishing or legitimate. 

Most published state-of-the-art in this section extract bespoke statistical features from the web page components, while others use a combination of existing features while adding a few of theirs. The studies by \cite{kumi2021malicious}, \cite{mohammad2012assessment}, \cite{rendall2020towards}, and \cite{smadi2018detection} proposed new statistical features to detect phishing web pages.

Specifically, \cite{kumi2021malicious} extracted eight features from web page content to a Classification Based on Association Rules Algorithm (CBA) to detect phishing web pages. These features include the number of special characters, sensitive words in a URL, and the entropy of the domain name. Evaluations of 700 phishing and 500 legitimate URLs yielded an accuracy of 95.8\%. To provide a more in-depth study of the type and number of features that can be extracted from a web page, \cite{mohammad2012assessment} developed a neural network model that automatically adapts the network structure to the ever-changing features needed to determine the status of a web page \cite{mohammad2014predicting}. The model had 17 features, including the presence of the $<$a$>$ tag and the ability to request URLs in a domain other than the one entered in the address bar. It was evaluated on 600 legitimate and 800 phishing websites, yielding an accuracy of 92.18\% in over 1000 epochs. 

\cite{rendall2020towards} proposed a multi-layered approach to detect phishing web pages based on 13 DNS-based features and multiple supervised machine learning algorithms. Their method, which included a sensors module (for data collection and preprocessing), a detection engine (with a supervised learning algorithm), and a triage module (for comparison with predefined thresholds), was tested on 17,244 legitimate and 7970 phishing domains, yielding an accuracy of 89\% using the SVM algorithm. Also, \cite{yerima2020high} proposed a phishing detection approach that took 30 static features from a web page's URL and HTML content and applied them to 1D convolutional neural networks. Experiments on 6,157 legitimate and 4,898 phishing websites achieved an accuracy of 98.2\% and an F1-score of 97.6\%. 

Additionally, some studies focused on detecting malicious domains, such as the study by \cite{maroofi2020comar}, who proposed COMAR (Classification of COmpromised versus MAliciously Registered Domains), a system capable of distinguishing compromised domains from malicious ones. COMAR uses 38 features to determine the state of a domain. The COMAR model was applied to 41,000 phishing URLs using a random forest classifier, yielding a 97\% accuracy. 

Some authors focused on evaluating existing features on various machine learning algorithms. \cite{moghimi2016new} used nine features derived from a subset of related studies \cite{zhang2012phishing}, \cite{lakshmi2012efficient}, and \cite{aburrous2010intelligent} and eight new attributes that indicate the relationship between the composition of a web page and its URL. The proposed features were derived using the Levenshtein distance algorithm \cite{yujian2007normalized}, which approximately identifies similarities in textual string patterns. Utilizing a dataset comprising 1448 phishing and 686 legitimate websites, the algorithm was classified using SVM, giving an accuracy of 99.14\%. \cite{chiew2019new} proposed the Hybrid Ensemble Feature Selection (HEFS) framework for machine learning-based phishing detection systems, which were implemented on a Random Forest classifier. HEFS is more computationally efficient and can determine the optimal number of features for a given dataset. For the UCI phishing dataset, their approach achieved an accuracy of 96.17\%. 

\cite{singh2015phishing} implemented the Adaline network and backpropagation algorithm with an SVM and neural network on 15 features based on a related study \cite{mohammad2012assessment}. The test was performed on a dataset of 179 phishing URLs and 179 legitimate URLs. The authors experimentally demonstrated that the Adaline network with SVM performed better than the backpropagation algorithm, with a prediction accuracy of 99.14\%.

\cite{aljofey2022effective} extracted features represented by URL character sequences which are combined and fed to train the XGBoost classifier. In particular, they extracted character-level Term Frequency-Inverse Document Frequency (TF-IDF) features from noisy parts of HTML and plaintext of 60,252 web pages to validate the proposed solution. This data contains 32,972 benign web pages and 27,280 phishing web pages. The proposed approach achieved an accuracy of 96.76\%.

One of the challenges of phishing detection using machine learning algorithms with handcrafted features is their difficulty in extrapolating to new data, and attackers might be aware of the specific features the model is trained on and can easily bypass it. To mitigate this challenge, \cite{smadi2018detection} proposed a neural network model that can adapt to the dynamic nature of phishing emails using reinforcement learning. The proposed model can handle zero-day phishing attacks and mitigate the problem of a limited dataset using an updated offline database. Their experiment yielded a high accuracy of 98.63\% for 50 features extracted from a dataset of 12,266 emails.

To explore the application of fuzzy logic in detecting phishing web pages, \cite{barraclough2013intelligent} built a model that combines fuzzy logic and neural networks to expose phishing in online transactions. The novelty combines 288 features from a user-behaviour profile, legitimate website rules, PhishTank, and pop-ups from emails. These attributes were trained using supervised learning and yielded an accuracy of 98.5\%. Although the authors stated that their model outperformed Netcraft \footnote{https://www.netcraft.com} and Cantina+ \cite{zhang2007cantina}, it is highly complex and resource-intensive. It is also heavily dependent on manual feature engineering, which is time-consuming.

Some researchers have shifted their focus from desktop to mobile web page security. Their research revealed that web experience on mobile phones differs functionally and structurally from desktop computers. These distinctions are primarily due to the mobile-specific features and capabilities to improve user experience. \citep{amrutkar2017detecting} proposed the KAYO. This binary classification algorithm distinguishes between legitimate and phishing mobile-specific web pages in real-time. The KAYO model employs machine learning techniques and heuristics derived from the page sources of mobile web pages, URLs, and mobile-specific facilities. Forty-four features were extracted and used in a binomial logistic regression machine-learning technique. KAYO achieved 89\% TPR, 8\% FPR, and 90\% accuracy.

Despite their high accuracy, the proposed algorithms that employ feature engineering techniques have a few limitations: (1) the inconvenient nature of manual feature engineering techniques, which require specialized domain knowledge to determine which features are useful for a given platform; and 2) the challenges faced by models built on predefined features when confronted with new data, as predefined features struggle to extrapolate to new data.

Furthermore, some of the proposed algorithms depend on the content of the associated legitimate web page, as phishers are expected to reproduce legitimate web page content and add handlers to save user data in their preferred repository \citep{moghimi2016new}. This may not be the case, as cybercriminals can easily design a website that looks like a legitimate website without completely replicating its content.

\subsection{Phishing Detection Using Automatic Feature selection on Deep Neural Networks}
Deep Neural Networks (DNN) use layers of stacked nonlinear projections to learn representations of multiple levels of abstraction. The state-of-the-art approaches reviewed in this section employ DNN algorithms to automatically learn the salient features in the web pages to detect phishing.

\cite{bahnsen2017classifying} proposed a phishing classification scheme that uses only the URLs of a web page as input and implements the model on an LSTM network. The results yielded a 98.7\% accuracy on a corpus of two million phishing and legitimate URLs. The authors compared their results with another model that implemented random forest (RF) on 14 lexical and statistical features extracted from the URLs. The LSTM model outperformed the latter model by 5\% across all metrics evaluated (accuracy, precision, recall and f1-score). Although RF had a faster runtime than the LSTM network, the authors demonstrated that the LSTM model does not require complete content analysis. 

Although the model proposed by \cite{wei2019deep} is very similar to the approach proposed by \cite{bahnsen2017classifying} in that they both use unprocessed URLs as input, the former employs novel word-embedding techniques for automatic feature representation. These features were then applied to convolutional filters to detect phishing URLs. Their experimental results on a dataset of 999,996 legitimate URLs and 523,970 phishing URLs yielded an accuracy of 86.6\%.

\cite{le2018urlnet} proposed URLNet, a deep learning model that concatenates character-and character‐level word embeddings as input and applies them to convolutional layers. URLNet was evaluated by its application to 5 million URLs, yielding an accuracy of 97.2\%.

The model proposed by \cite{ozcan2021hybrid} amalgamated manual and automatic feature extractions. Their hybrid deep learning model uses character embedding representation with 28 NLP features as input to a DNN + LSTM model. Their evaluations on a dataset of 37385 phishing URLs and 36,400 legitimate URLs yielded an accuracy of 98.79\% and an F1 score of 98.81\%.

\cite{zhang2021multiphish} developed MultiPhish, a phishing detection model that exploits a Variational autoencoder to fuse the text, image, and URL feature information of web pages with neural networks. The proposed model was applied to a dataset of 3887 phishing and 4259 legitimate instances, yielding an accuracy of 97.79\%. MultiPhish can detect web pages hosted in compromised domains by adding URL features. 

\cite{tang2021deep} extracted character-level features from the URL, which were then applied to a  gated recurrent unit (GRU) neural network to determine the maliciousness of the URL. Application of the model on 429, 125 legitimate URLs and 236,362 phishing URLs yielded an accuracy of 99.18\%.

A summary of these studies is provided in Table \ref{heuristics}.

{\footnotesize
\centering
\linespread{0.5}
\begin{longtable}[c]{p{0.2\textwidth}|p{0.3\textwidth}|p{0.26\textwidth}|p{0.2\textwidth}}
 \caption{Analysis of Phishing detection by Machine Learning Based Methods Using Manual and Automated Features.\label{heuristics}}\\
 \hline
 \hline
 \multicolumn{4}{c}{\textbf{Machine Learning Based Methods Using Manual Feature Engineering}} \\
\hline
\textbf{Authors}     & \textbf{Method} & \textbf{Dataset} 
&  \textbf{Performance} \\
 \hline
 \endfirsthead
 \hline
 \hline
\textbf{Authors}     & \textbf{Method} & \textbf{Dataset} 
&  \textbf{Performance}\\
 \hline
 \endhead
 \hline
 \endfoot
 \hline
 \hline
 \endlastfoot
   \cite{kumi2021malicious} & The authors applied 8 features to the CBA algorithm. The measure of the randomness factor in URLs was the most important feature. & 700 phishing and 500 legitimate websites. & 95.8\% accuracy\\ 
        \hline
         \cite{moghimi2016new} & The authors applied 9 features to the SVM algorithm. & 1448 phishing and 686 legitimate websites. & 99.14\% accuracy\\ 
        \hline
        \cite{singh2015phishing} & 15 features from the URL, HTML and networks were implemented on Adaline with SVM. & 79 phishing and 179 legitimate URLs. & 99.1\% accuracy. \\
        \hline
        \cite{mohammad2014predicting} & 17 features were implemented on a self-learning neural network. & 600 legitimate websites and 800 phishing websites. & 92.18\% accuracy. \\
        \hline
         \cite{chiew2019new} & 48 features were implemented on a Random Forest Classifier. & 2456 legitimate and phishing instances. & 96.17\% accuracy. \\
        \hline
       \cite{barraclough2013intelligent} & Used fuzzy logic to derive 288 features implemented on supervised machine learning algorithms. & Not stated. & 98.5\% accuracy on the Legitimate site rules. \\
        \hline
       \cite{rendall2020towards} & Used a multi-layered approach implemented on supervised machine learning algorithms. & 17,244 legitimate and 7970 phishing domains. & 89\% accuracy on the SVM algorithm. \\
         \hline
        \cite{yerima2020high} & This study applied 30 static features from the URL and HTML content of the web page to a 1D convolutional neural network. & 6,157 legitimate and 4,898 phishing websites. & 98.2\% accuracy and F1-score of 97.6\%.\\
          \hline
       \cite{smadi2018detection} & This study applied 50 features from the URL and HTML content of emails to a neural network with reinforcement learning. & 12,266 emails. & 98.63\% accuracy.\\
        \hline
        \cite{maroofi2020comar} & 38 features from the URL were applied to a random forest classifier to detect compromised domains. & 41,002 URLs. & 97\% accuracy.\\
        \hline
        \cite{aljofey2022effective} & Extracted character-level Term Frequency-Inverse Document Frequency (TF-IDF) features from noisy parts of HTML and plaintext. & 32,972 benign web pages and 27,280 phishing web pages & 96.76\% accuracy\\
        \hline
        \cite{amrutkar2017detecting} & 44 mobile-specific features from JavaScript, HTML and URL were applied to a logistics regression algorithm to detect mobile phishing pages. & 349,137 benign URLs and 5,231 malicious URLs. & 970\% accuracy.\\ 
         \hline
         \hline
 \multicolumn{4}{c}{\textbf{Machine Learning Based Methods Using Automatic Feature Selection}} \\
 \hline
        \cite{bahnsen2017classifying} & Applied an embedding of raw URLs on an LSTM network. & 2 million phishing and legitimate URL. & 98.7\% accuracy\\
        \hline
       \cite{wei2019deep} & Applied a word embedding of raw URLs on a Convolutional Neural network. & 999,996 legitimate URLs and 523,970 phishing URLs. & 86.6\% accuracy\\
          \hline
       \cite{ozcan2021hybrid} & This study proposed a hybrid deep learning model that concatenates NLP features with character embedding before applying them to a DNN+LSTM model. & 37385 phishing URLs and 36,400 legitimate URLs. & 98.79\% accuracy\\
        \hline
       \cite{le2018urlnet} & Convolutional Neural Networks were applied to both the characters and word embeddings of the URL String to learn the embedding in a jointly optimized framework. & 5 million URLs were used. & 97.2\% accuracy\\
        \hline
       \cite{zhang2021multiphish} & Developed MultiPhish that uses VAE to fuse the text, image, and URL features. & 5887
        phishing instances and 4259 legitimate instances. & 97.79\% accuracy\\
        \hline
        \cite{tang2021deep}  & Extracted character-level features from the URL, which were then applied to a  gated recurrent unit (GRU) neural network & 429, 125 legitimate URLs and 236,362 phishing URLs & 99.18\%
 \end{longtable}}
 
The majority of the above-mentioned state-of-the-art features are hand-crafted. The only work similar to ours is \cite{le2018urlnet}'s discussion of using a URL's character embedding on convolutions. Our solution leverages the character and word embedding layer to automatically learn vector representations of a web page's URL and HTML content to detect phishing attacks without requiring expert feature engineering. Additionally, as discussed above, our approach concatenates the output of the embedding layer before presenting it for convolutions, preserving the original HTML and URL content.

\section{The Proposed Model}\label{webphish_section}
In this section, we elaborate on the architecture of our proposed deep neural network model, WebPhish.

We define the problem of detecting phishing web pages using their URL and HTML content as a binary classification task of predicting two classes: \textit{legitimate} or \textit{phishing}. Given a dataset with \textit{R} web pages $\{(u_1,h_1, y_1)$, . . . , $(u_R,h_R, y_R)\}$, where ${u_r}$ and ${h_r}$ for \textit{r} = 1, . . . , \textit{R} represents the URL and HTML content of the $r$th web page from the dataset, and ${y_r\in \{0, 1\}}$ is its label. ${{y_r = 1}}$ corresponds to a phishing web page and ${{y_r = 0}}$ is a legitimate web page. This study aims to automatically obtain the URL representation, $u_r \mapsto U$, and HTML  $h_r \mapsto H$, where $U$ and $H$ represent the character and word embedding feature matrices of the raw URL and HTML content, respectively. Subsequently, the embedding matrices are employed to learn the discriminant model $f: X \mapsto Y $ used to classify a given web page, where $X$ is the concatenation of embedding matrices $U$ and $H$.

\begin{figure}[ht]
\includegraphics[width = 0.7\textwidth, height = 14.5cm]{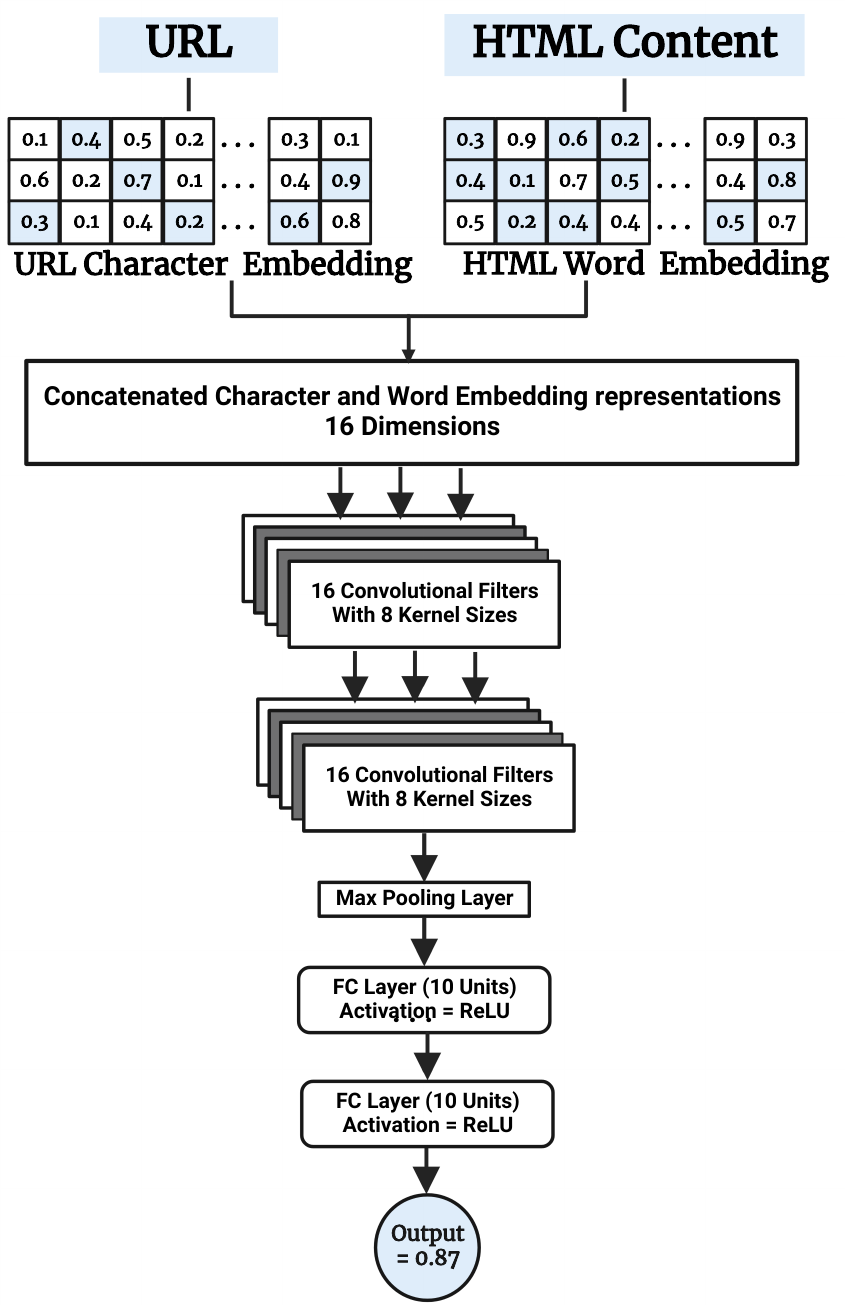}
\centering
    \caption{Overall Architecture of WebPhish.}
    \label{WebPhish_full}
\end{figure}

Figure \ref{WebPhish_full} provides an overview of the proposed model, WebPhish. WebPhish accepts a URL and HTML source code as input and then vectorizes the URL's characters and the words in the HTML source code using the Tokeniser utility class\footnote{\url{https://keras.io/api/preprocessing/text/}}. Each integer corresponds to a value in a dictionary that contains the entire corpus's keys, which are the vocabulary terms themselves. In the deep neural network, the tokenized characters and words are fed into the embedding layer as an array, and the weights are trained concurrently with the phishing detection process. The embedding layer, representing words and characters as dense vectors, captures the relationships between words and characters relevant to the task. The embedding layer then semantically maps related words such as "login" and "access" within the same embedding space.

Before connecting the embedding layers to the rest of the model, the URL embeddings are concatenated with the HTML embeddings to increase the corpus's vocabulary map. The concatenated features are then passed through convolution layers. CNNs were chosen because they are the best methods to satisfy our requirement to detect more intricate text patterns. We discuss comparisons of the CNN model with other candidates in Section \ref{tech_alternatives}. Finally, FC layers fuse the results from the convolution layer, and the sigmoid function is used to output the results.

The remainder of this section provides a detailed breakdown of each layer. 

\subsection{Input Layer}
An advantage of the proposed model is its ability to function directly with raw data. The first step in extracting the automated features in the web page is to create a dictionary of words that occurred in our dataset, with each word having its index. 

To create a dictionary of characters for the URLs, we consider each URL a character sequence and then identify all unique alphanumeric and special characters in the URLs in the dataset. 75 unique letters, numbers and special characters with a high frequency in the sequence set are selected to form a character dictionary. Finally, each URL character sequence in the sequence set is tokenised; a corresponding number replaces the original characters individually, such that a one-dimensional digital vector is obtained. This constitutes the character-level corpus of the URLs.

\begin{figure}[ht]
\includegraphics[width = 0.8\textwidth, height= 7.2cm]{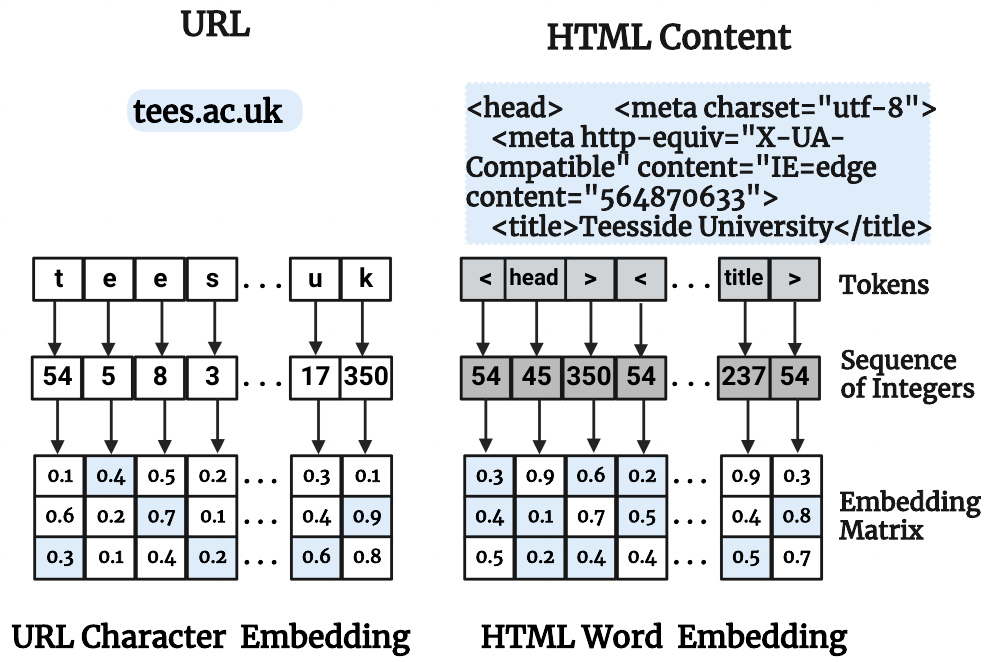}
\centering
    \caption{Configuration of the Embedding Layer in WebPhish }
   \label{embedding2}
\end{figure}

The construction of a word-level corpus of the HTML documents is similar to that of the character-level corpus of the URL. The difference is that HTML documents are segmented into word sequences instead of character sequences. To create a dictionary of words for the HTML documents, we split the content of the HTML document into individual words and treat all punctuation characters as separate tokens. For example, as shown in Figure \ref{embedding2}, <head>, is split into [\enquote{<}, \enquote{head}, \enquote{>} ]. The listed unique words create a dictionary in which every word becomes a feature. We obtained 321,009 unique words from the HTML content dataset. Finally, each HTML document word sequence in the sequence set is tokenized; a corresponding number replaces the original words individually to obtain the one-dimensional digital vector. This constitutes the word-level corpus of the HTML documents.

\subsection{Deep Neural Network}
The DNN has five layers, namely: (1) embedding; (2) concatenation; (3) convolution; (4) FC; and (5) output.

\subsubsection{Embedding Layer} Figure \ref{embedding2} shows the process in the embedding layer of the proposed model. In the embedding layer, the URL corpus's tokenized characters and the HTML document's tokenized words are automatically converted into feature vectors. Specifically, the raw data is transformed for each input using character embedding matrices into feature representations at the character level. The randomly initialized embedding matrices are gradually modified during training via backpropagation. Although some studies \cite{li2019stacking} have  used pre-trained Word2vec embeddings as features, other research has shown that learning the embeddings from task data optimized for a particular problem, maybe more efficient \cite{qi2018and}.

Consequently, we obtain the character embedding matrix of each URL $U$ $\in \mathbb{R}^{L_1 \times D} $ such that $L_1$ is the number of characters in each URL in the dataset, and $D$ is the embedding dimension. We experimentally set $D = 16$ and $L_1 = 180$. URLs over 180 characters were truncated from the 180th character, and URLs shorter than 180 were padded with <PAD> tokens until their lengths reached 180.

For the word embedding matrix of each HTML document $H$ $ \in \mathbb{R}^{L_2 \times D} $, $L_2$ is the number of words in each HTML document in the dataset, and $D$ is the embedding dimension. We experimentally set $D = 16$ and $L_1 = 2000$. 
 
\textit{Note:} Based on a manual analysis of our dataset, we determined that the longest URL had 156 characters, and the longest HTML document had approximately 1,898 words, excluding spaces. Therefore, we ensured the model did not lose critical web page information.

\subsubsection{Concatenation Layer} This layer concatenates the URL character and HTML word embedding matrices into a two-dimensional layer. The concatenation process adds two distinct characteristics to the model. First, it expands the vocabulary size of the embedding space, allowing it to exploit information from rare words and obtain a richer representation capable of capturing sub-word-level information. Second, concatenating the output from the embedding layer (without convolution filtering) preserves the original information of content that can also be used to detect phishing web pages.

\subsubsection{The Convolutional Neural Network}
The CNN consists of two major operations: convolution and pooling \cite{schmidhuber2015deep}. The convolution and pooling layers are stacked alternatively in the CNN framework until obtaining the high-level features on which an FC classification is performed \cite{lecun2015deep}. 

Taking the output from the concatenation layer with a shape  of (2180, 16), where 2180 is the sum of the length of the URL and HTML and 16 is the dimension, the convolution layer computes the output of neurons connected to the local regions in the input, each computing a dot product between their weights and a small region that they are connected to in the input volume. A ReLU activation function results in a matrix of [2180 x 32 x 8] as the model uses 32 filters and 8 kernels.  Max pooling is used for the feature map after the convolution operation to reduce network parameters and extract the most important features. Consequently, the POOL layer performs a downsampling operation along the spatial dimensions, resulting in the output of the convolutional layers being [1090 x 32 x 8], which is then passed to the following FC layers.

\subsubsection{Fully Connected Layer}
The model employs two FC layers that use the output of the convolution layers and merge the resulting features. Specifically, the two FC layers analyze the sequences from the CNN and max-Pooling layers and apply the ReLU activation in each FC layer.

\subsubsection{Output Layer}
The output layer, the final layer in our model, computes the model's result using the sigmoid function. This output layer, which follows the FC layers, compresses the model's result into a range of 0 to 1, according to the expression: $Q = \frac{1}{1 + {e}^-{^q}}$ given the probability of two classes: \textit{legitimate} or \textit{phishing}, where $q = (WR_t + b)$, \textit{W} and $b$ are the model parameters, and $R_t$ is the input at time step \textit{t}.

\section{Performance Evaluation}
This section elaborates on the research questions, dataset and evaluation metrics used to access WebPhish.

\subsection{Research Questions} \label{rq5}
We look at the following research questions:
\begin{itemize}
    \item \textbf{RQ1:} How accurate is WebPhish at detecting phishing pages compared to other deep learning-based state-of-the-art baselines?
    \item \textbf{RQ2:} How does WebPhish perform compared to hand-crafted features trained using SVM, Logistics Regression, and Random Forest?
    \item \textbf{RQ3:} What are the technical alternatives to the WebPhish model, and how effective are they?
    \item \textbf{RQ4:} How does the WebPhish model perform against other commonly used textual datasets for sentiment classification, such as the United States airline dataset?
\end{itemize}

To answer RQ1, in section \ref{compare}, we conduct experiments comparing WebPhish's performance to that of other baseline approaches on almost 46k phishing and benign webpages. To address RQ2, in section \ref{manual-features5}, we extract 31 popular features from our dataset's URLs and HTML source code and apply them to three classic machine learning algorithms: SVM, Logistics Regression, and Random Forest evaluating their performance against WebPhish. For RQ3, in section \ref{tech_alternatives}, we conduct a controlled experiment to assess performance when using alternative technical options for WebPhish, such as LSTM as the machine learning algorithm and URL or HTML as the only input. To address RQ4, in section \ref{airline}, we evaluate the model's accuracy using the US Airline dataset in order to demonstrate the model's flexibility on other textual datasets.

\subsection{Dataset}
To train our models, we used real-world datasets from Alexa.com for legitimate web pages and phishtank.com for phishing web pages to address the research questions above. The details are as follows: 

Benign web page dataset. We collated 22,687 benign web pages from the top-ranked Alexa list for this experiment. HTML documents were generated by coding a parser using the Beautiful Soup Library \cite{BeautifulSoup}. Beautiful Soup was chosen for two reasons: (1) it is functionally versatile and fast at parsing HTML content, and (2) it does not modify the HTML document object model composition when processing HTML documents. We created a parser that dynamically extracts the HTML source code for each web page from the final landing page. 

Phishing webpage dataset. Research has shown that machine-learning classiﬁers have diﬃculty coping with imbalanced train sets as they are sensitive to the proportions of the diﬀerent classes. These algorithms, in particular, tend to favour the majority class, resulting in misleading accuracy. As a result, we collated a balanced phishing set of 22,687 phishing  web pages from the study by \cite{korkmaz2020deep}. 

Subsequently, our final corpus contained a balanced dataset of \textbf{45,373} phishing and benign instances. Our datasets, URLs, and HTML for phishing and legitimate websites have been made available on\url{https://www.kaggle.com/datasets/guchiopara/look-before-you-leap}.



{\textbf{Note:} }
In the dataset, all the elements of an HTML document, such as text, hyperlinks, images, tables, lists, etc. and all parts of the URL, including the (subdomain, domain name, path, and query) were used when training the deep learning model. Also, we removed the prefix in URLs such as HTTP:// and HTTPS:// to prevent skewed results on different URL datasets.

\subsection{WebPhish Experimental Setup and Metrics}
To train WebPhish and its alternatives, a combination of hyperparameters was required. We used a grid search to determine our models' optimal number of CNN layers (1–3) and FC layers (1–3). Additionally, we determined the optimal optimization algorithm for the models (which varied between RMSProp and Adam) across a range of learning rates (from 0.0001 to 0.1).

\begin{table}
\small\addtolength{\tabcolsep}{-1pt}
\begin{minipage}[t]{0.9\textwidth}                      
  \begin{center}
\captionof{table}{WebPhish Hyperparameters}
\label{hyperparameters}
\begin{adjustbox}{center, width=\columnwidth}
\begin{tabular}{p{1.5in}|p{1.2in}|p{0.5in}}
      \hline
\textbf{Hyperparameters} & \textbf{Potential Choices} & \textbf{Selected} \\
\hline
\hline
Number of Conv1D layers & 1 -3 & 1 \\
        \hline
Number of FC Layers & 1 -3 & 2 \\
        \hline
Embedding Dimension & $4 - 32$ & 16 \\
        \hline
Optimizer & RMSProp and Adam & Adam \\
        \hline
        Learning rate   & 0.0001 - 0.1 & 0.0015\\
        \hline
        Number of Epochs & $5 - 30$ & 20\\
        \hline
        Batch Size   & $10 - 30$ & 20\\
        \hline
\end{tabular}
\end{adjustbox}
    \end{center}
\end{minipage}
\end{table}

Table \ref{hyperparameters} details the selected parameters we found gave the best performance on our dataset bearing in mind the unavoidable hardware limitation. We implemented all WebPhish variants in Python 3.5 on a Tensorflow 1.2.1 backend. We adjusted the batch size for training and testing the model to 20. The Adam optimizer \cite{kingma2015adam}, with a learning rate of 0.0015, was used to update the network weights. At the same time, we implemented binary cross-entropy to monitor the model's performance. The Early stopping technique \cite{prechelt1998early} was adopted to prevent overfitting of the training data. We conducted all WebPhish and baseline experiments on a Google Colaboratory environment with 12GB GDDR5 VRAM.

\subsection{Evaluation Metrics}\label{Metrics} We evaluated the performance of WebPhish using $Recall = \frac {(TP)}{(TP+FN)}$ and $Precision = \frac {(TP)}{(TP+FP)}$ where TP, FP and FN represent the numbers of True Positives, False Positives and False Negatives, respectively. The recall statistic is calculated as the number of correct results divided by the number of expected results. Finally, the Accuracy of WebPhish was determined using $
Accuracy = \frac {(TP+TN)}{(TP+TN+FP+FN)}$

We also used the receiver operating characteristic (ROC) curve and the precision-recall curve in our evaluation. The ROC curve is a probability curve, whereas the Precision-Recall curve depicts the trade-off between precision and recall for various threshold values. The receiver operating characteristic (ROC) curve is plotted against the false positive rate (FPR). In contrast, the precision-recall curve is plotted against the recall values. In addition, we also calculated the f1 score of the test from the precision and recall of the test using the formula $F1  score = \frac {Precision}{Recall}$.

Finally, we evaluated the machine learning model's performance using the K-fold cross-validation technique with k set to 5. Therefore, the data is divided into five folds and iterated five times. Cross-validation reduces problems such as overfitting and underfitting and indicates how the model will generalize to an independent dataset.

\section{Results}
This section discusses the experiments conducted to evaluate the proposed phishing detection method and their results in terms of each research question.

\subsection{Comparing WebPhish with State-of-the-Art Baselines (RQ 5.1)}\label{compare}
We compared WebPhish’s performance with existing state-of-the-art approaches for a comprehensive evaluation. These approaches include those by \cite{bahnsen2017classifying}, \cite{wei2019deep}, \cite{aljofey2022effective}, \cite{tang2021deep}, \cite{9207707}. The approach proposed by \cite{wei2019deep} is a DNN with multiple convolution layers that uses word tokens from a URL as input to determine the maliciousness of the associated web page. Moreover, the model proposed in \cite{bahnsen2017classifying} accepts the character sequence of a URL as input. It then uses LSTM neural networks to model the URL’s sequential dependencies to classify it as phishing or benign. 

HTMLPhish \cite{9207707} takes the character and word sequences from the HTML content of a web page as input and uses CNN to learn the semantic dependencies between them. The methods in \cite{bahnsen2017classifying}, and \cite{wei2019deep} studies were designed for phishing URL detection, and the study by \cite{9207707} was applied to the HTML contents. \cite{aljofey2022effective} extracted  URL character sequences features with textual content TF-IDF and hyperlink information features to create a feature vector applied to the machine learning models to detect phishing web pages.  To ensure an efficient comparison with our dataset, we applied only the TF-IDF features on XGBoost, as discussed in this chapter. \cite{tang2021deep} extracted character-level features from the URL, which were then applied to a  gated recurrent unit (GRU) neural network to determine the maliciousness of the URL.



\begin{table}
\centering
\small\addtolength{\tabcolsep}{-1pt}
\begin{minipage}[t]{0.8\textwidth}                   
  \begin{center}
    \caption{{Result of WebPhish and State-of-The-Art Baseline models}}
\label{baseline_result}
\begin{adjustbox}{center, width=0.9\textwidth}
\begin{tabular}{p{1.5in}|p{0.6in}|p{0.6in}|p{0.6in}|p{0.6in}|p{0.9in}}
        \hline
        {
\textbf{Models}} & {\textbf{Accuracy}} & {\textbf{Precision}} &  {\textbf{Recall}}  & {\textbf{F1 Score}} & {\textbf{Training time}} \\
\hline
\hline
{\textbf{WebPhish}} & {\textbf{0.981}} & { \textbf{0.982}} & {\textbf{0.981}} & {\textbf{0.981}} & {\textbf{491 Seconds}}\\
\cite{bahnsen2017classifying} & {0.953} & {0.963} & {0.942} & {0.952} & {258 Seconds}\\
\cite{wei2019deep}  & {0.873} & {0.804} & {0.982} & {0.884} & {533 Seconds}\\
\cite{9207707} & {0.942} & {0.909} & {0.980} & {0.943} & {338 Seconds}\\
        \cite{aljofey2022effective} & {0.951} & {0.951} & {0.953} & {0.951} & {135 Seconds}\\
        \cite{tang2021deep} & {0.960} & {0.948} & {0.971} & {0.960} & {600 Seconds}\\
        \hline
   \end{tabular}
\end{adjustbox}
    \end{center}
\end{minipage}
\end{table}

\textbf{Result:} Table \ref{baseline_result} summarises the results of WebPhish and the state-of-the-art baseline techniques. WebPhish exhibited the best performance across all metrics for the dataset. The results demonstrate that the concatenation of the character embedding of the URL and word embedding of the HTML content produces the highest accuracies by exploring the increased vocabulary and sequential patterns in the textual content. Notably, the methods proposed by \cite{bahnsen2017classifying} performed substantially better than those proposed by \cite{wei2019deep} using only word embeddings on URLs and by \cite{aljofey2022effective} using TF-IDF on the URLs.

\begin{figure*}[ht]
  \begin{subfigure}[b]{0.5\textwidth}
    \includegraphics[width = \textwidth]{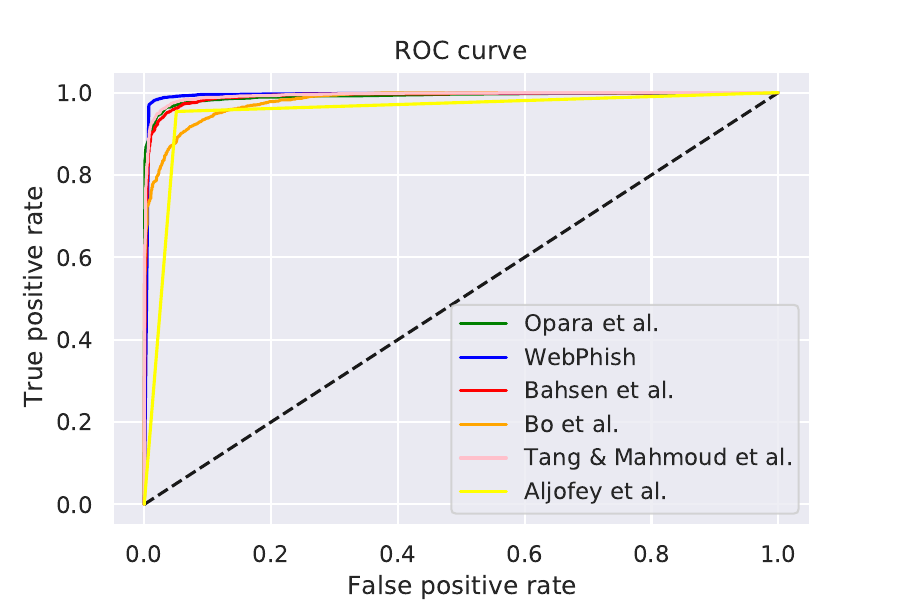}
    \caption{}
     \label{roc_baselines}
  \end{subfigure}
  \begin{subfigure}[b]{0.5\textwidth}
    \includegraphics[width = \textwidth]{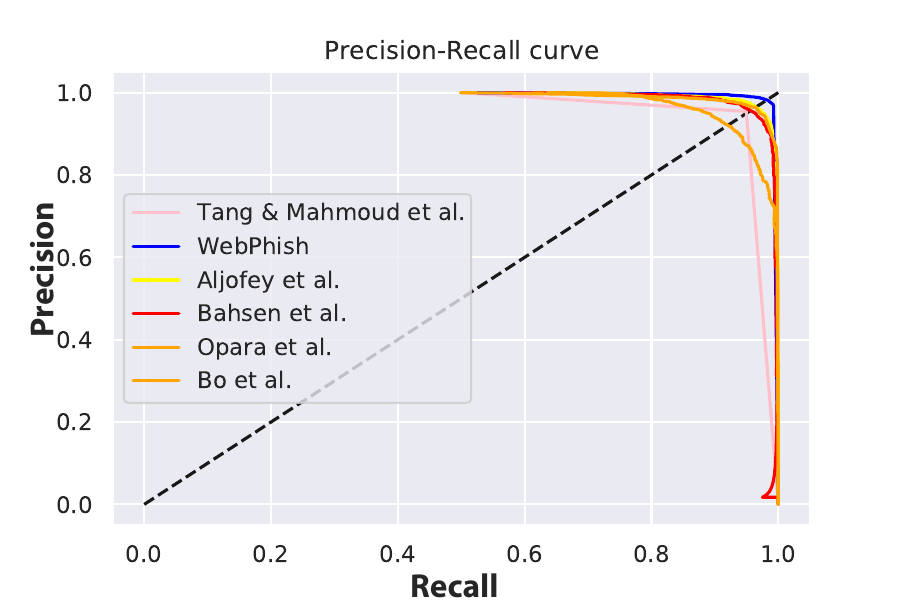}
    \caption{}
    \label{pr_baselines}
  \end{subfigure}
  \caption{{ROC and Precision-Recall Curves of WebPhish and its State-of-The-Art baselines}}
\end{figure*}

Additionally, we plot the Receiver Operating Characteristic (ROC) and the Precision-Recall curves for each phishing detection technique shown in Figures \ref{roc_baselines} and \ref{pr_baselines}. With decreasing FPR, we observe a growing gap between WebPhish and the baseline approaches, except for \cite{9207707}. Apart from WebPhish, the only other approach that achieves meaningful recall (TPR) at lower FPRs is \cite{9207707} (albeit this comes with a high computational cost). Even with low FPR values of 0.2, 0.3 and 0.4, which are required for operational deployment, WebPhish achieves a better recall than \cite{bahnsen2017classifying}.

The WebPhish model performed the best in terms of the F1 score, indicating that the model's overall performance is robust regarding precision and recall. In other words, the proposed methodology accurately detects phishing incidents and avoids misclassifying too many legitimate pages as phishing. This shows that the proposed phishing technique balances well between accuracy and recall.  

\subsubsection{Performance of the WebPhish Model in terms of Computational Time}
The main limitation of the proposed phishing detection model is that it is substantially more computationally intensive than other comparative models in terms of training time (see Table \ref{baseline_result}), with an average elapsed training time that is approximately two times higher than that of \cite{bahnsen2017classifying}. Overall, the proposed model is more computationally complex than other benchmarked methods, except for the model by \cite{wei2019deep}. In terms of runtime efficiency, the model requires 0.094s to evaluate whether a website is harmful based on its URL and HTML content. According to Fui-Hoon Nah et al., the acceptable wait time for web users to retrieve information is 2s. Therefore, when applied in real-time, the performance of our model does not negatively impact the website's usability.

\subsection{Comparing WebPhish with Hand-Crafted Features Trained on Simpler Baseline Models (RQ2)} \label{manual-features5}

This section compares WebPhish's deep neural network effect on phishing web page detection using raw URL and HTML content to simpler baseline models trained on hand-crafted features. We used three machine learning models: logistics regression, kernel SVM, and a random forest classifier. We chose these models because these traditional classifiers were commonly used in sequence detection systems \cite{mironczuk2018recent} and are relevant baselines to compare with WebPhish. We used \textbf{31} features detailed in Table \ref{features} culled from \cite{amrutkar2017detecting}, \cite{adebowale2019intelligent}, \cite{mohammad2012assessment}, and \cite{zhang2007cantina}.

{\small
\centering
\linespread{0.5}
\begin{longtable}[c]{p{0.3\textwidth}|p{0.7\textwidth}}
\caption{Extracted hand-crafted Features from Both URL and HTML Contents on Web Pages}
\label{features}\\
\hline
\hline
\multicolumn{2}{c}{\textbf{URL Features}} \\
\hline
Features &	Description \\
\hline
 \endfirsthead
 \hline
 \hline
Features &	Description \\
\hline
\endhead
 \hline
 \endfoot
  \endlastfoot
Number of misleading words in the URL, such as "login" and "bank." 
 &	Authors of phishing webpages often exploit the familiarity of users to a webpage \cite{amrutkar2017detecting} by including words in the URL that can mislead a user into believing that the phishing webpage is a legitimate webpage. Words such as login and bank are commonly used in the URL of the login webpage for benign websites that are highly prone to imitation.\\
\hline
Number and  \% of digits & Many phishing domain names are simply IP addresses of the machines that host them. To detect phishing presence, we calculated the number of digits in a URL and the percentage of digits in the hostname. \\
\hline
Number of forward slashes, question marks, dots, hyphens, underscores, equal signs, semicolons and ampersands & Phishers use some meaningless characters to confuse the victim. Therefore, they can also use some punctuation characters, especially “.”, ”;”, ”!”, ”\&”, ”\%”, etc. Increased value has more tendency to be a phishing webpage\\
\hline

Number and Presence of subdomains and two letter subdomains & Legitimate URLs typically have fewer subdomains; Also, phishers can use the subdomain names as if they were domain names. Furthermore, they can use several subdomains with names that are similar to the original ones. As a result, fewer subdomains increase the likelihood of a legitimate web page.\\
\hline
Length of the URL & Long URLs can be used by phishers to hide the suspicious part in the address bar. We averaged the lengths of the URLs in the dataset. Intuitively, the longer a URL is, the more likely it is to be a phishing URL.\\
\hline
\multicolumn{2}{c}{\textbf{HTML Features}} \\
\hline
 Number of NoScript & Intuitively, a benign webpage developer will include more NoScript in the code to ensure a good experience for even the most security-savvy user. \\
 \hline
Presence and  number of embedded JavaScript & A webpage that includes embedded JavaScript loads faster than a page that must refer to external code. According to previous research \cite{fette2006learning}, legitimate websites will have more embedded JavaScript than phishing websites. \\
\hline
Presence and number of external and internal JavaScript & Legitimate web pages use more internal and external JavaScript for advertising and analytics than malicious web pages\\
\hline
Number and Presence of internal and external links & Legitimate web pages have more internal links pointing to subdirectories on the website and fewer external links pointing to other legitimate websites. This feature also improves the user experience of the website, which is considered by legitimate website creators. Phishing webpages usually use external resources like their phishing targets to enrich their content in order to deceive users that the page is legal, resulting in very few internal links and many external links in a phishing webpage.\\
\hline
 Presence of iframes & IFrame is an HTML tag that displays an additional webpage within the one that is currently displayed. Phishing website developers use the "iframe" tag and make it invisible, i.e., without frame borders. As a result, the presence of iframes indicates a phishing website. This feature counts the number of strings that contain "iframe" in a script. \\
\hline
Presence of and number of images & Legitimate web pages contain more images for aesthetic and improved user experience, whereas malicious web pages contain fewer images.\\
\hline
Percentage of white spaces in the HTML content & Legitimate websites will have more white space between the code to ensure a good experience even for a security-savvy user. As a result, the more whitespaces there are in the HTML content, the more likely the website is legitimate. \\
\hline
        \hline
     \end{longtable}}

\subsubsection{URL Features} For the hand-crafted features extracted from the URL, research has shown that phishing web pages developers frequently exploit an Internet user's familiarity with a website \cite{dhamija2006phishing} by adding terms to the URL that may trick a user into thinking that somehow the malicious website is the real website. Widely used terms to access genuine websites, like admin and account, make them particularly vulnerable to imitation. Therefore, the creator of a phishing website would intuitively use ambiguous terms at the URL's start. Therefore, including those terms in the URL is regarded as a feature. Many malicious domain names are hosting systems IP addresses \cite{fette2007learning}, \cite{mcgrath2008behind}. We counted the combination of numbers in a URL and the percentage of numbers in the hostname as a feature. Also, phishers create several subdomains to include tricky terms, for example, PayPal as a subdomain. This could make phishing URLs longer \cite{mcgrath2008behind}. Therefore, we included the URL length, if the URL consists of a subdomain, the number of sub-domains, and the number of dots as features. Furthermore, the number of punctuation marks, such as semicolons, hyphens, and underscores, are included in our URL feature set.

\subsubsection{HTML Features} For the HTML feature set, variables such as the number of white spaces, the presence of internal and external links, and the number and presence of images were extracted because of their relevance when differentiating between a phishing and legitimate web page.

A binary classifier is taught using the extracted features, which are provided when the features are collected. We empirically set the number of trees as 70 for the random forest classifier, the penalty for the logistics regression as \textit{L1}, and the kernel bias function (RBF) of the non-linear SVM as 50.0.

\begin{table}[ht]
\centering
\small\addtolength{\tabcolsep}{-1pt}
\begin{minipage}[t]{0.8\textwidth}                   
  \begin{center}
    \caption{{Result of WebPhish and hand-crafted Features on Traditional Models}}
\label{manual_result}
\begin{adjustbox}{center, width=0.9\textwidth}
\begin{tabular}{p{2.7in}|p{0.6in}|p{0.6in}|p{0.6in}|p{0.6in}|p{0.9in}}
        \hline
{\textbf{Models} }    & {\textbf{Accuracy}} & {\textbf{Precision}} 
& { \textbf{Recall}}  & {\textbf{F1 Score}}   & {\textbf{Training time}} \\
        \hline
        \hline
{\textbf{WebPhish}} & {\textbf{0.981}} & {\textbf{0.982}} & {\textbf{0.981}} & {\textbf{0.981}} & {\textbf{491 Seconds}}\\
    {Kernel SVM + hand-crafted Features} &  {0.873} & {0.836} & {0.924} & {0.878} & {125 Seconds}   \\
       {Logistics Regression + hand-crafted Features} &  {0.884}    
& {0.865} & {0.908} & {0.885} & {65 Seconds}\\
       { Random Forest Classifier + hand-crafted Features} & {0.945} & {0.942} &{0.948} & {0.945} & {90 Seconds}\\
        \hline
   \end{tabular}
\end{adjustbox}
    \end{center}
\end{minipage}
\end{table}

\begin{figure*}[ht]
  \begin{subfigure}[b]{0.5\textwidth}
    \includegraphics[width = \textwidth]{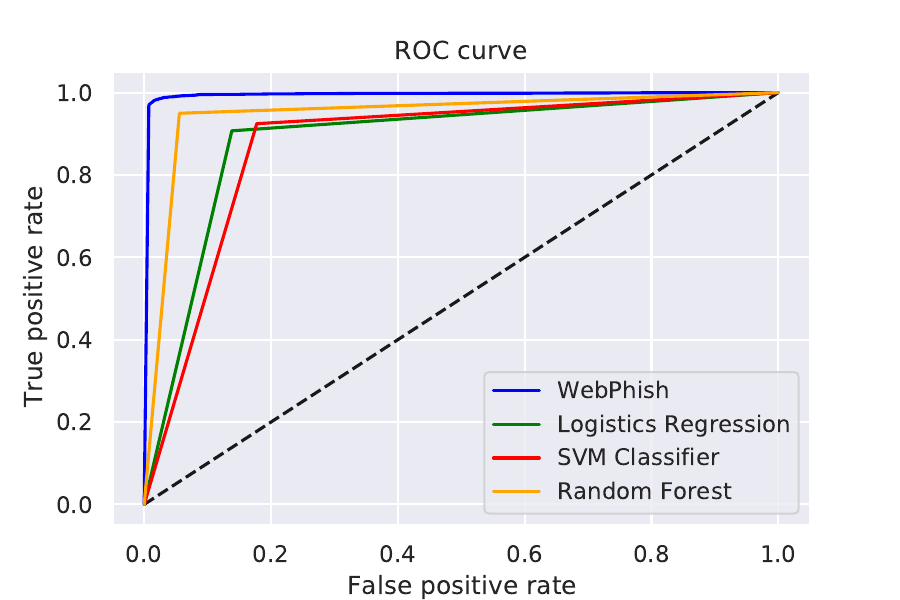}
    \caption{}
     \label{roc_manual}
  \end{subfigure}
  \begin{subfigure}[b]{0.5\textwidth}
    \includegraphics[width = \textwidth]{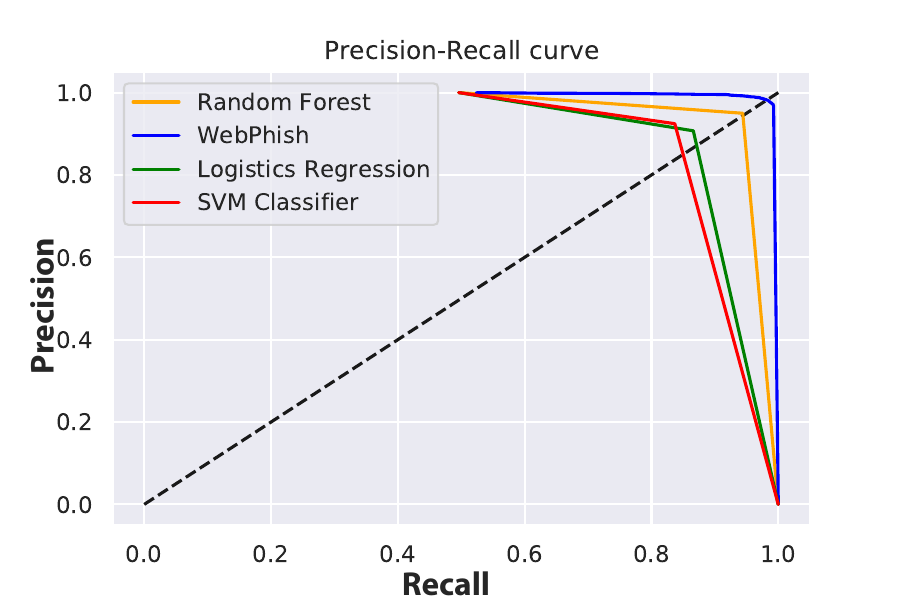}
    \caption{}
    \label{pr_manual}
  \end{subfigure}
  \caption{{ROC and Precision-Recall Curves of WebPhish and hand-crafted Features on Traditional Models}}
\end{figure*}

\textbf{Result:} Table \ref{manual_result} shows the precision, recall, and F1 score of WebPhish compared with the traditional machine learning classifiers. Figures \ref{roc_manual} and \ref{pr_manual} show the traditional machine learning classifier ROCs. WebPhish outperforms all state-of-the-art techniques (with an improvement in accuracy of at least 2\%) in all categories and metrics. 

The WebPhish classifier correctly predicted approximately 98\% of the 2,252 instances of phishing in the test set. In contrast, the SVM, logistic regression, and RF classifiers correctly predicted 88\%, 87\%, and 94\% of the phishing instances, respectively. The outputs of the traditional machine learning classifiers show the limitations of hand-crafted features. This highlights the importance of the temporal robustness of our proposed method.

In addition, the ROC curve in Figure \ref{roc_manual} shows that the RF classifier yields better results than the logistic regression and SVM classifiers, with its curve being further from the hyperplane.  


\begin{table*}[ht]\footnotesize\renewcommand{\arraystretch}{1.2}\addtolength{\tabcolsep}{-1.7pt}
\caption{{Confusion Matrix of WebPhish and hand-crafted Features on Traditional Models}} 
\label{confusion_matrix2}
\begin{center}
\begin{tabular}{|c|c|c|c|c|c|c|c|c|c|c|c|c|c|c|c|c|c|c|c|}
\hline
 & \multicolumn{2}{|c|}{\bfseries {WebPhish}}  & \multicolumn{2}{|c|}{\bfseries {Random Forest Classifier}} & \multicolumn{2}{|c|}{\bfseries {Kernel SVM}} & \multicolumn{2}{|c|}{\bfseries {Logistics Regression}}\\
\hline\noalign{\smallskip}
\cline{1-1} \cline{2-3}\cline{4-5}\cline{6-7} \cline{8-9} 
         \multicolumn{1}{|c|}{{ Legitimate}} & \multicolumn{1}{|c|}{{2237}} & {50}  &  \multicolumn{1}{|c|}{{2156}} & {131}  & \multicolumn{1}{|c|}{{2090}} & {197}  & \multicolumn{1}{|c|}{{1968}} & {319}         \\    
            \hline
     \multicolumn{1}{|c|}{ {Phishing}} & \multicolumn{1}{|c|}{{49}} & {2202}   &  \multicolumn{1}{|c|}{{117}} & {2134}  & \multicolumn{1}{|c|}{{162}} & {2089}  & \multicolumn{1}{|c|}{{207}} & {2044}    \\
\cline{1-1} \cline{2-3}\cline{4-5}\cline{6-7} \cline{8-9} 
                   \multicolumn{1}{|c|}{} & \multicolumn{1}{|c|}{{Legitimate}} & {Phishing}  &    \multicolumn{1}{|c|}{{Legitimate}} & {Phishing}  & \multicolumn{1}{|c|}{{Legitimate}} & {Phishing} &  \multicolumn{1}{|c|}{{Legitimate}} & {Phishing}     \\
\cline{1-1} \cline{2-3}\cline{4-5}\cline{6-7} \cline{8-9} 
\end{tabular}
    
\end{center}
\end{table*}


Furthermore, Table \ref{confusion_matrix2} above is the confusion matrix for the proposed model (WebPhish) and the traditional machine learning classifiers on the dataset. With 4538 instances in the test set, the classifier correctly predicted "phishing" 2202 times and "legitimate" 2237 times for the WebPhish model yielding a precision of 98.1\%.

In the scenario where the HTML content is cloned, and the URL has been spoofed, the model using these features will leverage the URL and HTML features to determine the maliciousness of the web page. For example, if the HTML content has been cloned word by word, and the URL is spoofed by using only the IP address of the website, then the HTML features, including the length of the page, number of images and external links, will not suffice as they will be same in both the cloned and legitimate web page. However, the model will rely on URL characteristics such as the number and percentage of digits in the URL to identify the maliciousness of the web page.

\subsection{Alternative Technical options for WebPhish (RQ3)}\label{tech_alternatives}
In addition to the proposed end-to-end framework, which combines URL embedding at the character level with HTML content, we evaluated alternative technical options for implementing WebPhish. Here we examine the following technical alternatives:

\begin{itemize}
    \item {Option1 (LSTM): This deep learning model learns URL features and HTML content representations using RNNs, particularly \textbf{LSTM}. The character embedding matrix is applied to the LSTM layer, whose output is concatenated and sent to the dense layers. Then, the sigmoid layer outputs the classification results.}
    \item {Option2 (URL): This CNN model is trained exclusively on the \textbf{URL}. The embedding layer's character embedding matrix is applied to the CNN and max-pooling layers, which are then passed to the dense layers. The results are output via the sigmoid layer.}
    \item {Option3 (HTML): This is a CNN model trained exclusively on the \textbf{HTML content}. The embedding layer's word embedding matrix is applied to the convolution and max-pooling layers, and the results are passed to the dense layers. The sigmoid layer outputs the results.}
\end{itemize}

\begin{figure*}[ht]
  \begin{subfigure}[b]{0.5\textwidth}
    \includegraphics[width = \textwidth]{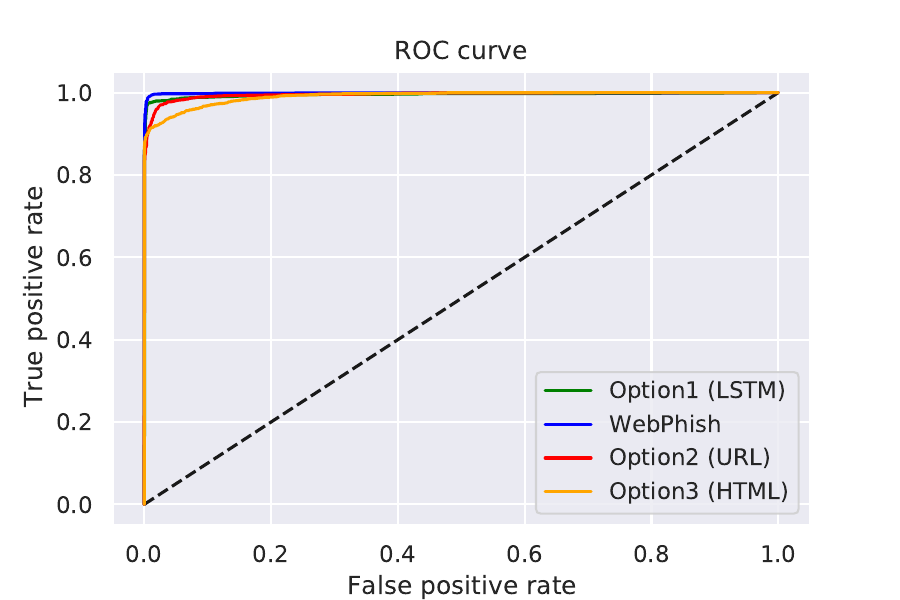}
    \caption{}
     \label{roc_alternative}
  \end{subfigure}
  \begin{subfigure}[b]{0.5\textwidth}
    \includegraphics[width = \textwidth]{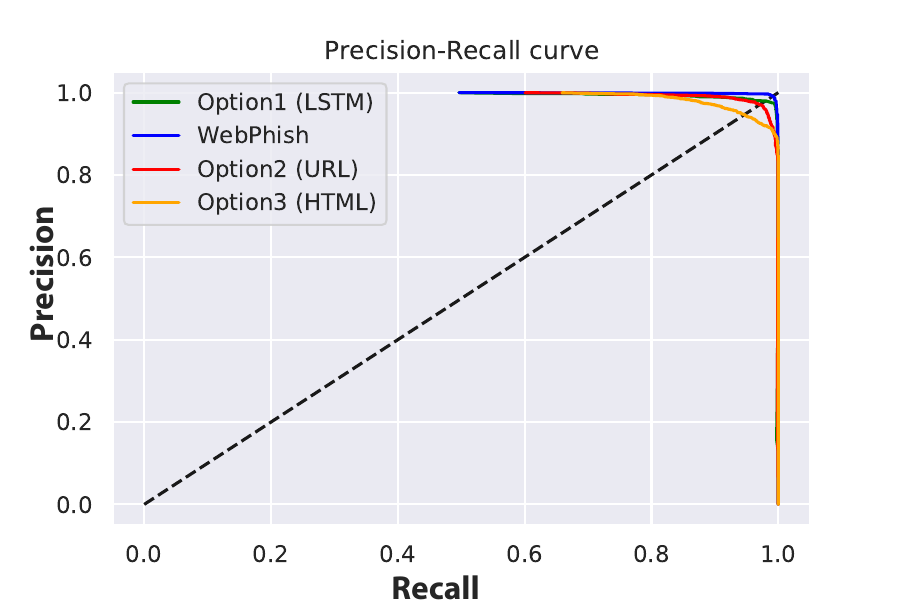}
    \caption{}
    \label{pr_alternative}
  \end{subfigure}
  \caption{{ROC and Precision-Recall Curves of WebPhish and its Variants}}
\end{figure*}

{\textbf{Result:} In Figure \ref{roc_alternative} and Figure \ref{pr_alternative}, we show the ROC curve of WebPhish and its alternative options. Amongst the alternative options, Option3 (HTML) was the least performing across all evaluated metrics. This outcome is because phishing web pages, especially those hosted on compromised websites, are known to systematically copy the legitimate web page source code in other to blend in effortlessly. In contrast, Option1 (LSTM) performed more closely to the proposed model, albeit having a higher training time. Furthermore, a random selection of false positive URLs showed the presence of known phishing words such as "login" and "account," which typically occur in phishing URLs. However, the model can accurately classify a high percentage of the legitimate URLs because, upon closer examination of our dataset, we discovered that some words more frequently occur in phishing web pages than legitimate ones and vice versa. For example, in our dataset, the word "login" appeared 99.8\% more often in phishing than in legitimate URLs. Consequently, the model generates a prediction based on the semantic meaning of the highlighted text and its co-occurrence on the page.}

\begin{table}
\centering
\small\addtolength{\tabcolsep}{-1pt}
\begin{minipage}[t]{0.8\textwidth}                   
  \begin{center}
    \caption{{Result of WebPhish and its Alternative Options}}
\label{Webphish_alternative}
\begin{adjustbox}{center, width=0.9\textwidth}
\begin{tabular}{p{1in}|p{0.6in}|p{0.6in}|p{0.6in}|p{0.6in}|p{0.9in}}
        \hline
{\textbf{Models}}     & {\textbf{Accuracy}} & {\textbf{Precision}} 
&  {\textbf{Recall}}  & {\textbf{F1 Score}}   & {\textbf{Training time}} \\
        \hline
        \hline
{\textbf{WebPhish}} & {\textbf{0.981}} & {\textbf{0.982}} & {\textbf{0.981}} & {\textbf{0.981}} & {\textbf{491 Seconds}}\\
        {Option2 (URL)} & {0.959} & {0.956} & {0.964} & {0.960} & {193 Seconds} \\
        {Option3 (HTML)} & {0.965} & {0.970} & {0.961} & {0.965} & {533 Seconds}\\
{Option1 (LSTM)} & {0.954} & {0.965} & {0.943} & {0.954} & {614 Seconds}\\
         \hline
        \hline
   \end{tabular}
\end{adjustbox}
    \end{center}
\end{minipage}
\end{table}

In general, Table \ref{Webphish_alternative} indicates that WebPhish significantly outperforms the three options: Option1 (LSTM), Option2 (URL) and Option3 (HTML). WebPhish achieved an average of 98.1\% across its precision, F1 score, and recall metrics on the dataset. WebPhish leverages the strength of the alternatives and consistently produces better results while capturing local and temporal patterns in the data. Furthermore, the precision, recall, and F1 score for WebPhish are well-balanced as their values are similar. This indicates that WebPhish can detect phishing web pages accurately when implemented in the real world. Furthermore, WebPhish achieved the lowest FPR of 2\% as it leverages the strengths of both the URL and HTML input to consistently produce better results while capturing local and temporal patterns in the data. This indicates that the proposed model can learn the characters and words in the dataset and their associated semantic meanings to identify phishing web pages.

\subsection{Application on Sentiment classification (RQ4)}\label{airline}
We evaluated our model's performance on the publicly available US airline dataset to demonstrate its adaptability to other textual datasets. The US airline dataset was sourced from CrowdFlower's Kaggle databases\footnote{https://www.kaggle.com/crowdflower/twitter-airline-sentiment}. There are 14,640 tweets in this data set. American Airlines, United Airlines, US Airways, Southwest Airlines, Delta Airlines, and Virgin Airlines are among the airlines that have used these tweets. Sentiment classification techniques can aid researchers and decision-makers at airlines in better comprehending their customers' feelings, opinions, and satisfaction.

We fed WebPhish the actual text tweeted by customers and the airline sentiment confidence associated with it. The airline sentiment confidence metric is a numeric value that indicates the degree of certainty associated with categorizing a tweet as neutral, positive, or negative.

\subsubsection{Implementation}
Using the evaluation metrics detailed in Section \ref{Metrics}, we applied the WebPhish model to the US Airline dataset to evaluate the model's ability to differentiate between positive and negative emotions in text. 

Note: as the US Airline dataset has been publicly available since 2015, previous studies have been conducted on classifying its sentiments. We compared the performance of WebPhish on the US Airline dataset with the following studies: In \cite{rustam2019tweets}, the authors applied a VC to classify tweets according to their emotions. The VC is based on logistic regression and SGDC and uses a soft voting mechanism to obtain the final prediction. 

In addition, using TF-IDF as a feature extraction mechanism, the authors implemented a phrase-level analysis on seven classification algorithms: decision tree, RF, SVM, Gaussian Naive Bayes, AdaBoost, logistic regression, gradient boosting, and the VC. We also compared WebPhish with the system proposed by \cite{tanggated}. Using Doc2Vec embeddings on the US Airline dataset, \cite{tanggated} explored a bi-directional GRU network for sentiment analysis of Twitter data directed at US airlines. They fed a trained word vector to the bi-directional GRU network using a skip-gram model, which was initialized by the existing GloVe model \cite{socher2013recursive}.

   \begin{table}[!htb]
\centering
   \small\addtolength{\tabcolsep}{-1pt}
\begin{minipage}[t]{0.9\textwidth}  
\begin{center}
    \caption{Result of WebPhish and State-of-the-art models on the US Airline Dataset Classifying 3 classes: Positive, Negative and Neutral}
    \label{us_airline1}
    \begin{adjustbox}{center, width=0.9\textwidth}
      \centering
     \begin{tabular}{l|l|l|l|l c|}
\hline
       \multirow{2}{*} & \textbf{Corpus Size} & \textbf{No. of Classes} & \textbf{Algorithm} & \textbf{Accuracy in \%} \\
       \hline
       \hline
        &  & & Decision Tree & 63 \\ 
        & & & Random Forest & 75.8 \\
        & &   & SVM & 78.5 \\
            & & & Gaussian Naive Bayes & 43.8 \\
       \cite{rustam2019tweets}  & 14,640 & 3 Classes & AdaBoost & 74.6 \\
             & & & Logistic Regression & 78.7 \\
             & & & Gradient Boosting  & 73.4 \\
             & & & Decision Tree Classifier & 68.6 \\
             & & & Voting Classifier & 79.2 \\
             \hline
             \cite{tanggated}  & 14,640 & 3 Classes & Gated GRU & 74 \\
              \hline
             \textbf{WebPhish}  & \textbf{14,640} & \textbf{3 Classes} & \textbf{DCNN} & \textbf{80} \\
          \hline
     \end{tabular}
     \end{adjustbox}
       \end{center}
\end{minipage}
   \end{table}

\begin{table*}
\centering
\small\addtolength{\tabcolsep}{-1pt}
\begin{minipage}[t]{0.9\textwidth}                      
  \begin{center}
  \caption{Result of WebPhish and State-of-the-art models on the US Airline Dataset Classifying 2 classes: Positive and Negative}
    \label{us_airline2}
\begin{adjustbox}{center, width=0.9\textwidth}
\begin{tabular}{p{1.5in}|p{0.8in}|p{0.8in}|p{0.8in}|p{1in} p{0.5in}}
      \hline
    & \textbf{Corpus Size} & \textbf{No. Classes} & \textbf{Algorithm} & \textbf{Accuracy in \%} \\
        \hline
        \hline
             \cite{naseem2019dice}  & 11,542 & 2 Classes & BiLSTM & 93.6 \\
              \hline
              \cite{khan2020big}  & 11,542 & 2 Classes & Deep RNN & 93 \\
              \hline
            \textbf{WebPhish}  & \textbf{11,542} & \textbf{2 Classes} & \textbf{DNN} & \textbf{94.1} \\
\hline
\end{tabular}
\end{adjustbox}
  \end{center}
\end{minipage}
\end{table*}

We compared our model with the DICET method proposed in \cite{naseem2019dice}. DICET is an automated text pre-processor fed into a Bidirectional LSTM with attention to detecting Twitter sentiment analysis. Also, \cite{khan2020big} proposed a sentiment analysis model that extracts relevant features from the US airline dataset using a Hadoop cluster. The obtained features are fed into a deep RNN network to perform the classification, providing two classes: positive and negative.

\cite{naseem2019dice} and \cite{khan2020big} removed the neutral class from the US Airline dataset during experimentation, thereby reducing the dataset to 11,541 tweets. Conversely, \cite{rustam2019tweets} and \cite{tanggated} experimented with the entire corpus of the US Airlines dataset of 14,640 instances. Consequently, WebPhish was trained and evaluated on full and abridged datasets to ensure fairness.

Tables \ref{us_airline1} and \ref{us_airline2} detail the result of WebPhish and other state-of-the-art techniques on the US airline dataset. WebPhish indicates a significant improvement in efficiency compared with current state-of-the-art techniques trained on the  US airline Twitter datasets.

WebPhish, with an accuracy of 94\%, works better than other models for the US airlines dataset. We can effectively conclude that our proposed model is a robust solution and applies to other text classification fields beyond social engineering. 

\section{Discussion}
\subsection{Why does WebPhish outperform the baselines?}
In this study, the proposed model WebPhish, effectively tackled the detection of phishing web pages using only the URL and HTML raw content on a deep learning model.

The advantages of WebPhish over the baseline models include: 
\begin{enumerate}
    \item The approach presented in this study leverages the dual input approach using both the raw URL and HTML content to detect phishing attacks. For example, in the scenario where the URL is spoofed, the HTML content is relied upon to provide a convincing result on the maliciousness of the web page. Also, unlike existing techniques, our solution does not rely on manual feature engineering. It exploits the URL and HTML content's character and word embedding matrices to detect a phishing attack without expert feature engineering.
    \item Choice of semantic level to use in the embedding layer: Existing research influenced the semantics level chosen for the embedding layer \cite{opara2019htmlphish}. Word-level embeddings are limited to the training data's unique dictionary of words. Consequently, it cannot generate useful feature representations for newly introduced words in the test data. According to our analysis, character-level embeddings outperform word-level embeddings. With a limited number of existing characters in languages, the character embedding layer is less likely to encounter characters not present in the test dataset. As a result, it is simple to apply to test data. However, in the phishing detection task, the approaches proposed by \cite{wei2019deep} and HTMLPhish struggled to distinguish information for scenarios in which phishing URLs and HTML documents attempt to impersonate benign web pages using obfuscation techniques. This is because convolution filters produce similar output from a sequence of characters with the same spelling or only minor differences. Consequently, CNNs based on URLs or HTML alone are insufficient for obtaining detailed structural information from a web page. Hence, we considered the concatenation of both web page component embeddings.

    In summary, WebPhish uses the URL's character embedding matrices and the HTML's word embedding matrices to accommodate unseen words in the test data, yielding better results than the other variants.
    \item  Introducing the concatenation layer before feeding the input features to the convolution layers: Unlike previous work \cite{le2018urlnet}, which concatenates the outputs of convolution layers, WebPhish combines the outputs of embedding layers. Concatenating the output of the embedding layer (without convolutional filtering) preserves the original information regarding the content useable for detecting malicious web pages, as demonstrated in our experiments.

    In WebPhish, before connecting the embedding layers to the rest of the model, URL embeddings are concatenated with HTML embeddings, increasing the corpus's vocabulary map. The concatenated features are then passed through convolution layers. Consequently, WebPhish can learn patterns based on the URL characters and HTML word embeddings.
\end{enumerate}

\subsection{How does WebPhish perform on cloned websites?}

WebPhish's main objective is to accurately determine the maliciousness of a given website using its HTML and URL. When its HTML content and URL are spoofed precisely in the same format, WebPhish uses the character and word level characteristics of the HTML and URL to distinguish these spoofed websites from the benign ones. 

For example, suppose the URL is spoofed by misspelling the characters in the links or by using non-Latin characters to make homographic URLs. Because this proposed approach is trained on a dataset of multi-language websites and uses character embeddings to learn the text's semantic, lexical, and syntactic ambiguities, the model will detect these URL spoofing methods. When an Internet user clicks on a shortened URL, the full URL is disclosed before the web page is opened. As a result, the model can still evaluate the full URL to determine its maliciousness. 

If the HTML content is cloned word for word, the proposed method will depend on the URL content to determine the maliciousness of the website. To summarize, the proposed approach uses both the contents of URLs and HTML to remain effective in the face of a phishing attack.

{\subsection{The Importance of the Number of Layers in the DNN on the Phishing Detection Accuracy}}\label{ablation}
{To demonstrate the importance of the DNN layers in WebPhish in detecting phishing, we examine FC layers and CNN layers' effect on the proposed model efficiency. }

Table \ref{fc_layer} shows the results of our evaluations on the effect of the FC layers. Intuitively, we expect that more FC layers will increase the model's accuracy. However, our analysis found that the proposed model’s configuration of 2 FC layers gave our task's best performance on the dataset, while an extra layer did not improve the accuracy. The proposed model achieved an accuracy of 97.9\%, 98.1\%, and 98\% with 1, 2, and 3 FC layers.

\begin{table}
\centering
\small\addtolength{\tabcolsep}{-1pt}
\begin{minipage}[t]{\columnwidth}                      
  \begin{center}
\captionof{table}{{The Impact of The FC Layers}}
  \label{fc_layer}
\begin{adjustbox}{center, width=\columnwidth}
\begin{tabular}{p{1.7in}|p{0.6in}|p{0.9in}}
    \hline
 {\textbf{Models}}    & {\textbf{Accuracy}} & {\textbf{Training time}} \\
\hline
\hline
{1 FC Layers} & {0.979} & {475 Seconds} \\
        \hline
{Proposed Model (2 FC Layers)} & {0.981} & {491 Seconds} \\
        \hline
{3 FC Layers} & {0.980} & {492 Seconds}  \\
        \hline
        \end{tabular}
\end{adjustbox}
    \end{center}
\end{minipage}
\end{table}
\begin{table}
\centering
\small\addtolength{\tabcolsep}{-1pt}
\begin{minipage}[t]{\columnwidth}                      
  \begin{center}
\captionof{table}{{The Impact of The Convolutional layers}}
  \label{CNN_filters}
\begin{adjustbox}{center, width=\columnwidth}
\begin{tabular}{p{2.3in}|p{0.6in}|p{0.9in}}
      \hline
     {\textbf{Models}} & {\textbf{Accuracy}} & {\textbf{Training time}} \\
\hline
\hline
{1 Convolutional Layer} & {0.980} & {480 Seconds}  \\
        \hline
{ Proposed Model (2 Convolutional Layers)} & {0.981} & {491 seconds} \\
        \hline
{3 Convolutional layers} & {0.980} & {493 Seconds} \\
        \hline
        \end{tabular}
\end{adjustbox}
    \end{center}
\end{minipage}
\end{table}

\begin{table}
\centering
\small\addtolength{\tabcolsep}{-1pt}
\begin{minipage}[t]{0.9\textwidth}                      
  \begin{center}
\captionof{table}{{The Impact of The Embedding Layer}}
  \label{emb_layer}
\begin{adjustbox}{center, width=0.9\textwidth}
\begin{tabular}{p{2.5in}|p{0.6in}|p{0.9in}}
        \hline
{\textbf{Models} }    & {\textbf{Accuracy}}  & {\textbf{Training time}} \\
\hline
\hline
{Proposed Model} & {0.981}  & {491 Seconds} \\
        \hline
{Proposed Model without Embedding layer} & {0.941}  & {268 Seconds} \\
        \hline
        \end{tabular}
\end{adjustbox}
    \end{center}
\end{minipage}
\end{table}

Table \ref{CNN_filters} shows the effect of the number of CNN layers in the model. We found that 2 CNN layers gave the best balance of training time of 491 seconds and accuracy of 98.1\%. When using 1 and 3 CNN layers, WebPhish can achieve 98\% and 98.1\% accuracy and a training time of 475 seconds and 492 seconds, respectively.

Furthermore, we demonstrated the importance of the embedding layer in the DNN model in phishing detection. We analyzed this by checking the performance of WebPhish on the dataset when the embedding layer was replaced with hand-crafted features from the URL and HTML characteristics on the CNN and FC layers. Table \ref{features} lists the hand-crafted features. Although the hand-crafted features' training time is shorter than the embedding features, a 4\% drop  is observed across all metrics in Table \ref{emb_layer}. This highlights the character embedding matrix’s importance when analyzing textual content in the URL and HTML. It also demonstrates that the tedious hand-crafted feature engineering process can overlook salient characteristics that differentiate a phishing web page from a legitimate one.

\subsection{Limitations of WebPhish}
While we argue that WebPhish can detect zero-day phishing attacks that contain known phishing HTML and URL content, if the zero-day phishing attack involves a rare manipulation of the web page content, WebPhish may miss it. Additionally, due to the rare manipulation of web page content on web pages, the model will need to be retrained to remain relevant. Additionally, as the proposed model is not visual based, the model cannot classify web pages whose DOM is predominantly embedded images or flashes.

\section{Conclusion}
In this paper, we propose a technique for detecting web page phishing. We used a DNN, more precisely Convolutional Neural Networks, to capture the semantic relationships inherent in raw URL and HTML content while initiating automatic feature extraction via character and word embedding techniques. The proposed method overcomes some limitations associated with existing approaches that rely on hand-crafted features or one raw web page component and enables straightforward extrapolation to new data. Additionally, as it takes the proposed model 0.094 seconds to evaluate whether a website is harmful, therefore, when applied in real-time, the performance of our model will not negatively impact the website’s usability when deployed in the real world. The next step will be to examine our model’s performance against other types of phishing attacks, such as spear phishing.

\bibliography{main}

\end{document}